\documentclass[12pt]{article}
\newlength{\pubnumber} 

\catcode`\@=11
\@addtoreset{equation}{section}

\def\section{\@startsection{section}{1}{\z@}{3.5ex plus 1ex minus .2ex}
 {2.3ex plus .2ex}{\large\bf}}
\def\subsection{\@startsection{subsection}{2}{\z@}{2.3ex plus .2ex}
 {2.3ex plus .2ex}{\bf}}

\usepackage{latexsym}

\renewenvironment{thebibliography}[1]
        {\begin{list}{[\arabic{enumi}]}
        {\usecounter{enumi}\setlength{\parsep}{0pt}
         \setlength{\itemsep}{0pt}
         \settowidth
        {\labelwidth}{[{#1}]}\sloppy}}{\end{list}}

\begin{document}

\begin{titlepage}
\samepage{
\setcounter{page}{0}
\rightline{January 2008}
\vfill
\begin{center}
    {\Large \bf New Regulators for Quantum Field Theories\\
    with Compactified Extra Dimensions\\
        {\it --- Part II:  ~Ultraviolet Finiteness and~~~~~\\ 
   Effective Field Theory Implementation ---}\\}
\vfill
   {\large Sky Bauman\footnote{ E-mail address:
      bauman@physics.arizona.edu} $\,$and$\,$ Keith
      R. Dienes\footnote{ E-mail address:  dienes@physics.arizona.edu}
      \\}
\vspace{.10in}
 {\it  Department of Physics, University of Arizona, Tucson, AZ  85721
 USA\\}
\end{center}
\vfill
\begin{abstract}
  {\rm  In a previous companion paper [arXiv:0712.3532], 
        we proposed two new regulators 
        for quantum field theories in spacetimes with compactified extra dimensions.
        Unlike most other regulators which have been used in
        the extra-dimension literature, these regulators are 
        specifically designed to respect the original
        higher-dimensional Lorentz and gauge symmetries that exist prior to
        compactification, and not merely the four-dimensional symmetries which
        remain afterward.
        In this paper, we use these regulators in order to develop a 
        method for extracting ultraviolet-finite results from one-loop
        calculations.  This method  also allows us 
        to derive Wilsonian effective field theories for 
        Kaluza-Klein modes at different energy scales.  
        Our method operates by ensuring that 
        divergent corrections to parameters describing the physics of the excited 
        Kaluza-Klein modes are absorbed into
        the corresponding parameters for zero modes, thereby eliminating the need to
        introduce independent counterterms for parameters characterizing
        different Kaluza-Klein modes.  
        Our effective field theories can therefore simplify calculations 
        involving Kaluza-Klein modes,
        and be compared directly to potential 
        experimental results emerging from collider data.}
\end{abstract}
\vfill
\smallskip}
\end{titlepage}

\setcounter{footnote}{0}

\def\beq{\begin{equation}}
\def\eeq{\end{equation}}
\def\beqn{\begin{eqnarray}}
\def\eeqn{\end{eqnarray}}
\def\half{{\textstyle{1\over 2}}}
\def\third{{\textstyle{1\over 3}}}
\def\quarter{{\textstyle{1\over 4}}}

\def\calO{{\cal O}}
\def\calE{{\cal E}}
\def\calT{{\cal T}}
\def\calM{{\cal M}}
\def\calF{{\cal F}}
\def\calS{{\cal S}}
\def\calY{{\cal Y}}
\def\calV{{\cal V}}
\def\ibar{{\overline{\imath}}}
\def\chibar{{\overline{\chi}}}
\def\ttwo{{\vartheta_2}}
\def\tthree{{\vartheta_3}}
\def\tfour{{\vartheta_4}}
\def\ttwob{{\overline{\vartheta}_2}}
\def\tthreeb{{\overline{\vartheta}_3}}
\def\tfourb{{\overline{\vartheta}_4}}

\def\qbar{{\overline{q}}}
\def\mm{{\tilde m}}
\def\nn{{\tilde n}}
\def\rep#1{{\bf {#1}}}
\def\ie{{\it i.e.}\/}
\def\eg{{\it e.g.}\/}

\newcommand{\newc}{\newcommand}
\newc{\gsim}{\lower.7ex\hbox{$\;\stackrel{\textstyle>}{\sim}\;$}}
\newc{\lsim}{\lower.7ex\hbox{$\;\stackrel{\textstyle<}{\sim}\;$}}

\hyphenation{su-per-sym-met-ric non-su-per-sym-met-ric}
\hyphenation{space-time-super-sym-met-ric}
\hyphenation{mod-u-lar mod-u-lar--in-var-i-ant}


\def\inbar{\,\vrule height1.5ex width.4pt depth0pt}

\def\IC{\relax\hbox{$\inbar\kern-.3em{\rm C}$}}
\def\IQ{\relax\hbox{$\inbar\kern-.3em{\rm Q}$}}
\def\IR{\relax{\rm I\kern-.18em R}}
 \font\cmss=cmss10 \font\cmsss=cmss10 at 7pt
\def\IZ{\relax\ifmmode\mathchoice
 {\hbox{\cmss Z\kern-.4em Z}}{\hbox{\cmss Z\kern-.4em Z}}
 {\lower.9pt\hbox{\cmsss Z\kern-.4em Z}} {\lower1.2pt\hbox{\cmsss
 Z\kern-.4em Z}}\else{\cmss Z\kern-.4em Z}\fi}

\long\def\@caption#1[#2]#3{\par\addcontentsline{\csname
  ext@#1\endcsname}{#1}{\protect\numberline{\csname
  the#1\endcsname}{\ignorespaces #2}}\begingroup \small
  \@parboxrestore \@makecaption{\csname
  fnum@#1\endcsname}{\ignorespaces #3}\par \endgroup}
\catcode`@=12

\input epsf
\section{Introduction
\label{intro}
}
\setcounter{footnote}{0}

If all goes according to plan, the Large Hadron Collider (LHC) 
will uncover exciting new phenomena at the TeV scale.
These phenomena are likely to hold clues pertaining
to some of the  most pressing current mysteries of particle physics, 
including the nature of electroweak symmetry breaking and the origin of
the stability of the energy scale at which this occurs.
Indeed, through such discoveries, 
data from the LHC is likely to change the paradigm of 
high-energy physics, eventually leading to a new ``Standard Model'' 
for the next generation of particle physic(ist)s. 

Of course, if we subscribe to the belief that the truly fundamental
energy scales of physics are unreachably high 
(\eg, at or near $M_{\rm Planck}\approx 10^{19}$~GeV, or
at least significantly above the electroweak scale),  
then this new ``Standard Model'' will be at best 
yet another effective field theory (EFT), valid only within a well-prescribed
energy range.
Interpreting this data-produced effective Lagrangian 
will then require comparisons to the EFT's which
can be derived from various potential theoretical models of possible new physics.
For example, weak-scale supersymmetry (SUSY) is widely considered to be a 
compelling candidate for new physics, and most phenomenological studies 
of weak-scale SUSY focus on specific EFT's
(\eg, the Minimal Supersymmetric Standard Model) in which the supersymmetry is
broken but in which the origin of this breaking is not included.

Extra spacetime dimensions are also leading candidates for new physics beyond
the current Standard Model.
However, while there has been considerable work analyzing the cumulative effects
that the corresponding towers of Kaluza-Klein (KK) states might have
on ordinary four-dimensional physics, 
there have been almost no studies concerned with the 
EFT's of the towers of excited KK modes themselves.
Analyses which do exist are qualitative,
focus on special interactions (\eg, brane kinetic terms), or contain
special implicit assumptions.

Yet there are general EFT questions which might be asked in this context.
For example, if there exists 
a single extra flat dimension
compactified on a circle of radius $R$, then the masses of the corresponding KK modes 
can be expected to follow the well-known relation $m_n^2=m_0^2 + n^2/R^2$.  
Likewise, the couplings of 
these modes will all be equal:  $\lambda_{n,n',...}=\lambda_{0,0,...}\delta_{n+n'+...}$.
These relations are nothing but the reflection of the higher-dimensional
Lorentz invariance which holds in the ultraviolet (UV) limit, and such patterns will
be taken as strong evidence in judging whether newly discovered particles 
are indeed KK states.
However, as one passes to lower energies ({\it e.g.}\/, through a Wilsonian renormalization
group analysis), these masses and couplings are subject to radiative corrections.
As a result, we expect that these simple mass and coupling relations will be deformed as 
the heavy KK states are integrated out of the spectrum.
Indeed, at relatively low energies, the spectrum of low-lying KK modes may
be significantly distorted relative to our na\"\i ve tree-level expectations,
and this can potentially be important for experimental searches for 
(and the identification/interpretation of) such states.  

The goal of this paper is to develop methods of deriving and analyzing the EFT's
of such KK towers as functions of energy scale.
Indeed, if extra dimensions are ultimately observed at the LHC through the discovery
of KK resonances, it will be important to understand the radiative corrections to
the masses and couplings of such states since this information will ultimately feed into 
precision calculations
of their cross-sections and decay rates. 
However, aside from potential experimental consequences, analyzing 
the EFT's of towers of KK resonances as functions of the energy scale
is also interesting from a purely theoretical perspective, since 
this provides the only systematic way of understanding what happens as 
extra dimensions are ``integrated'' out in passing from a higher-dimensional UV limit
to a four-dimensional infrared limit.  
 
One fundamental obstacle to performing such a renormalization-group 
analysis of the KK spectrum has been that general
techniques for regularizing loop effects in KK theories were not
known.  While quantum-mechanical regulators exist 
which preserve the four-dimensional
symmetries (such as Lorentz invariance and gauge invariance) 
which remain after compactification,
such regulators are sufficient only for radiative calculations
of the physics of the zero modes.
By contrast, calculations of the {\it excited}\/ KK modes
will require techniques which preserve the full set of {\it higher-dimensional} 
symmetries.  While there has been a small literature concentrating on radiative
corrections in KK theories (see, \eg, Refs.~[2-18]), relatively few approaches actually satisfy
this latter requirement. 

In Ref.~\cite{paper1a}, we developed two new regulators for quantum
field theories in spacetimes with compactified extra dimensions.
We refer to these regulators as the ``extended hard cutoff'' (EHC)
and ``extended dimensional regularization'' (EDR).
Although based on traditional four-dimensional regulators,
the key new feature of these higher-dimensional regulators is that
they are specifically designed to handle mixed spacetimes in which some
dimensions are infinitely large  and others are compactified.
Moreover, unlike most other regulators which have been used in
the extra-dimension literature, these regulators are designed to
respect the original
higher-dimensional Lorentz and gauge symmetries that exist prior to
compactification, and not merely the four-dimensional symmetries which
remain afterward.  

By respecting the full higher-dimensional symmetries, the regulators of
Ref.~\cite{paper1a} avoid the introduction of spurious terms which would not have been
easy to disentangle from the physical effects of compactification.
Moreover, by preserving the physics associated with higher-dimensional
symmetries, they maintain the associated Ward identities.
For example, in a gauge-invariant theory,
analogues of the Ward-Takahashi identity should hold not only for
the usual zero-mode (four-dimensional) photons, but for all
excited Kaluza-Klein photons as well.
It is the regulators in Ref.~\cite{paper1a} which preserve such identities
for the excited KK modes as well as the zero modes.

In this paper, we will extend the techniques in Ref.~\cite{paper1a} in two directions.
\begin{itemize}
\item  First, we shall show how the regulators of Ref.~\cite{paper1a} can be used
in order to extract ultraviolet-finite results from one-loop calculations.
Our method operates by ensuring that
        divergent corrections to parameters describing the physics of the excited
        Kaluza-Klein modes are absorbed into
        the corresponding parameters for zero modes, thereby eliminating the need to
        introduce independent counterterms for parameters characterizing
        different Kaluza-Klein modes.
\item  Second, we shall show how these finite results can be used in order to construct
effective field theories (EFT's) for towers of Kaluza-Klein (KK)
modes.  Our EFT approach will therefore provide a framework for comparing an
effective Lagrangian extracted from LHC data to higher-dimensional
theoretical models. 
Additionally, as we shall discuss, our EFT's will carry special 
advantages for calculations of loop effects in experiments involving excited KK modes.
\end{itemize}

In this paper, we shall follow a Wilsonian approach
towards deriving our EFT's.
Specifically, we shall employ the regulators of Ref.~\cite{paper1a} 
to calculate the masses and couplings of the KK states
as functions of a Wilsonian renormalization group scale.
In other words, we shall explicitly integrate out heavy KK states
above a given scale $\mu$,
and observe how the parameters describing the remaining light
(but nevertheless excited) KK states are affected as a function of
the scale $\mu$.
One key observation will be essential to this analysis:
Although the masses and couplings of individual KK states can
be expected to experience strong divergences,
the {\it relative differences}\/ of these parameters between excited 
KK modes and the zero mode are physical observables and thus can be 
expected to remain finite and regulator-independent.
Thus, if the parameters describing the zero modes at a given energy scale
are assumed to be determined from experiment, then these finite differences
can be used to obtain the parameters describing
all of the other excited KK states at this scale.
We thus obtain all the parameters needed to define EFT's 
describing the tower of KK states as functions
of energy scale. 

Although the techniques presented here are more general than
most previously existing methods, our analysis in this paper will
be restricted in certain significant ways.
First, our procedures will apply to calculations in
theories for which the compactification space is a smooth manifold
rather than an orbifold.
As such, we will not be considering the effects of extra terms which
might arise at singularities of the compactification space, such
as brane kinetic terms.  Moreover, as discussed above, although the differences 
between KK parameters should be finite regardless of any perturbative expansion,
this paper will focus exclusively on one-loop calculations.
Finally, although our techniques can easily be generalized,
for concreteness we shall primarily consider the case
of a single extra dimension compactified on a circle.

This paper is organized as follows.  In Sect.~\ref{rel}, we shall
show how to use the regulators from Ref.~\cite{paper1a} in order
to extract finite, regulator-independent predictions in 
KK theories at one-loop order.
Specifically, we shall provide
a general procedure for
deriving finite, regulator-independent expressions for differences
between renormalized KK parameters.  Then, in 
Sect.~\ref{examples}, we shall provide two explicit examples
illustrating how this procedure is implemented.
In Sect.~\ref{EFT},
we shall then demonstrate how to obtain
regulator-independent Wilsonian EFT's from these
differences.  Specifically, we shall show how to calculate Wilsonian
evolutions of EFT parameters with respect to the energy scale. 
Our conclusions can be found in Sect.~\ref{conclude}.

This paper is the second in a two-part series, and relies on the
results from an earlier companion paper~\cite{paper1a}.  As such, we shall
assume complete familiarity with the methods of Ref.~\cite{paper1a}, and
shall not review results which can be found there.

\section{Ultraviolet Finiteness
\label{rel}}
\setcounter{footnote}{0}

In this section, we shall provide a general procedure
for calculating finite,
regulator-independent corrections to differences between parameters
characterizing excited modes and zero modes in KK towers. 
As indicated in the Introduction, we shall rely on
the methods developed in Ref.~\cite{paper1a}, and we shall assume 
that the reader is familiar with these techniques.
Sect.~\ref{examples} will then provide two explicit examples 
illustrating how this procedure is implemented.

\begin{figure}[ht]
\centerline{
   \epsfxsize 3.0 truein \epsfbox {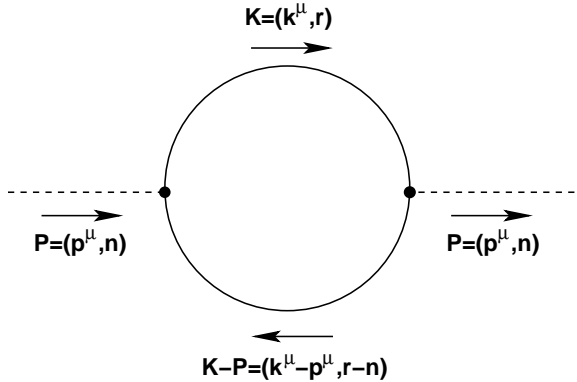} }
\caption{A generic one-loop diagram, as in Ref.~\protect\cite{paper1a}:  
   an external Kaluza-Klein particle (dotted line)  with four-momentum
   $p^\mu$ and Kaluza-Klein index $n$ interacts with a tower of
   Kaluza-Klein particles (solid lines) of bare mass $M$.} 
\label{fig1}
\end{figure}

We begin by considering a generic one-loop
diagram of the form shown in Fig.~\ref{fig1} in which an external
particle with four-momentum $p^\mu$ and mode number $n$ interacts with
a tower of KK particles  of bare (five-dimensional) mass $M$.
For example, the mass of the $r^{\rm th}$ KK mode in this tower
is given by $m_r^2=M^2 + r^2/R^2$ if the extra dimension is compactified
on a circle.
Enforcing 5D momentum
conservation at the vertices (as appropriate for compactification on a
circle) then leads to a one-loop expression of the general form
\beq
           L_n (p) ~=~ i \int_{0}^{1}dx \, \sum_{r} \, f_n (p,r,x)
\label{kgenform}
\eeq
where $x$ is a Feynman parameter and where $f_n$ represents an
appropriate four-dimensional loop momentum integral.

In general, such an expression will diverge badly.  
Meaningful algebraic manipulations are therefore only possible in the
presence of a regulator.  In this case, there are two possible 
sources of divergence:  the four-momentum integral $f_n$,
and the internal KK summation $\sum_r$.
Both must therefore be regulated, and, as discussed in the Introduction,
we need to utilize regulators which preserve the full {\it five-dimensional}\/
symmetries which exist prior to compactification.
These include not only five-dimensional Lorentz symmetry, but also
five-dimensional gauge symmetry when appropriate.
The fact that five-dimensional symmetries must be preserved
implies that we must somehow {\it correlate}\/ the regulator for the
four-momentum integral with the regulator for the KK summation
so that they are both imposed and lifted together.

In Ref.~\cite{paper1a}, two such regulator procedures were introduced.
In our extended hard cutoff (EHC) procedure, the four-momentum integral
$f_n$ is regulated through the introduction of a hard cutoff, while 
our extended dimensional regularization (EDR) procedure utilizes 
a generalization of ordinary 't~Hooft-Veltman dimensional regularization for $f_n$.
In either case, the KK summation is regulated through the introduction
of a hard cutoff $\Lambda$, and all appropriate five-dimensional
symmetries are protected through the introduction of
strict relations between this cutoff $\Lambda$
and the regulator parameters involved in regulating $f_n$.
These relations are given in Ref.~\cite{paper1a}. 
Note that while our EDR procedure is completely general, preserving both five-dimensional
Lorentz and gauge symmetries, our EHC regulator preserves only Lorentz symmetries
and thus is suitable for theories without gauge symmetries.    
In either case, the net result is that the general 
expression in Eq.~(\ref{kgenform}) then takes the regulated form
\beq
           L_n (p) ~=~ i \int_{0}^{1}dx \, 
          \sum_{r= -\Lambda R+xn}^{\Lambda R+xn}\,  f_n (p,r,x)~
\label{kgenform2}
\eeq
where we now understand $f_n$ to denote an 
appropriate {\it regulated}\/ four-momentum integral,
and where the particular form of the KK limits   
is explained in Ref.~\cite{paper1a}, with the notation $\sum_{r=a}^b$ denoting
a summation over integer values of $r$ within the range $a\leq r\leq b$ (even if $a$
and $b$ are not themselves integers).
We shall also assume that $\Lambda R$ can
be treated as an integer;  in the $\Lambda\to\infty$ limit as our cutoff
is removed, this assumption will not affect our final results.
Of course, it is understood when writing expressions such 
as Eq.~(\ref{kgenform2}) that we are to take the limit $\Lambda\to\infty$
at the end of the calculation (along with a simultaneous, correlated
removal of the regulator within the four-momentum integral). 

Our goal is to obtain finite, regulator-independent expressions for differences
such as $L_n-L_0$ for $n\not=0$.
In order to do this, we begin by utilizing an identity discussed 
in Ref.~\cite{paper1a}.
Specifically, for any $n\not=0$, we can perform a series of variable
substitutions in order to write
\beq
           \int_0^1 dx\, \sum_{r=-\Lambda R+xn}^{\Lambda R+xn}  ~=~
         {1\over |n|} \sum_{j=0}^{|n|-1} \, \int_0^{1} d\hat u \,
         \sum_{\hat r = -\Lambda R + 1}^{\Lambda R}~,
\label{kgenid}
\eeq
where
\beq
           \hat u ~\equiv~ x|n|-j
\label{u_hat}
\eeq
and
\beq
             \hat r ~\equiv~ {\rm sign}(n) r - j~.
\label{r_hat}
\eeq
This identity serves to render the KK summation cutoffs
independent of the Feynman parameter.
Thus, rewriting $L_n$ for $n\not=0$ in this way and dropping 
the hats from $\hat u$ and $\hat r$, we have
\beq
      -i(L_n - L_0) ~=~  
     \frac{1}{|n|}\sum_{j = 0}^{|n| - 1} \int_{0}^{1}du\,  
           \sum_{r = -\Lambda R + 1}^{\Lambda R} 
                     \, f_n (r, u,j) ~-~  
     \int_{0}^{1}dx \, \sum_{r = -\Lambda R}^{\Lambda R}\, f_0(r,x)~.
\label{prediff}
\eeq
Relabeling $x\to u$ in the second term and using the fact that $f_0$ is
$j$-independent to join the integrands, 
we thus have
\beqn
      -i(L_n - L_0)  &=& 
          \sum_{r = -\Lambda R}^{\Lambda R}
         \frac{1}{|n|}\sum_{j = 0}^{|n| - 1}\int_{0}^{1}du \,
         \biggl\lbrack f_n (r,u,j) -f_0(r,u)\biggr\rbrack\nonumber\\
         && ~~~~~~~~~~~-~ \frac{1}{|n|}\sum_{j = 0}^{|n| - 1}
                 \int_{0}^{1}du \, f_n (-\Lambda R,u,j)~.
\label{diff}
\eeqn
Note that we have dropped the four-momentum $p$ from the expressions
for  $L_0$ and $L_n$, since these expressions are presumed to be 
evaluated after appropriate renormalization conditions 
have been applied.

As discussed in the Introduction, 
physical observables such as the {\it relative}\/ masses and couplings 
between different KK states must remain finite
even though the masses and couplings for individual KK states 
might accrue divergent radiative corrections.
Indeed, relative differences such as these are originally finite 
at tree-level (modulo potential effects due to classical rescalings),
and are also finite to all orders in the UV limit (or equivalently the
$R\to\infty$ limit) where five-dimensional
Lorentz invariance is restored and the effects of compactification become
irrelevant.\footnote{This is not true in orbifold theories, due to the possible
          presence of brane kinetic terms.}
Since the divergence structure of the theory should
not be altered by changing $R$,
we expect such relative differences to remain finite regardless of
the radius or effective energy scale.
As a result, we expect that expressions such as those in Eq.~(\ref{diff})
should be finite either exactly as written, or 
with $L_0$ and $L_n$ replaced with those sub-expressions
within $L_0$ and $L_n$ which are responsible for renormalizing observables.
(For example, if the external KK particle in Fig.~\ref{fig1} is a KK photon
carrying Lorentz index $\mu$,
then the relevant sub-expression would consist of those terms $\tilde L_n^{\mu\nu}$
within the full $L_n^{\mu\nu}$ which are proportional to the metric $g^{\mu\nu}$
and which therefore renormalize the masses of the excited KK photons.)
We shall assume that 
our generic expressions $L_n$ consist of only such terms in what follows.
Note that since we are restricting our attention to one-loop 
diagrams, radiative corrections to relative KK parameters 
will indeed correspond to 
linear differences of the forms $L_n-L_0$.

Even though Eq.~(\ref{diff}) is finite,  
our goal is to write $L_n-L_0$ in
a manifestly finite, regulator-independent fashion.
In other words, we seek to be able to write differences such as $L_n-L_0$
in the analogous form
\beq
       -i(L_n - L_0)~=~   \sum_{r =
        -\infty}^{\infty}\,\frac{1}{|n|}\,\sum_{j = 0}^{|n| - 1} \,
        \int_{0}^{1}du \, \biggl\lbrack \alpha_n (r,u,j) - \alpha_0
        (r,u) \biggr\rbrack ~+~ \Delta_n ~,
\label{nocut}
\eeq
where the functions $\alpha_0$, $\alpha_n$, and $\Delta_n$ are 
are each manifestly finite and regulator-independent. 
However, comparing Eqs.~(\ref{diff}) and (\ref{nocut}),
we see that we are nearly there.
Indeed, looking at Eq.~(\ref{diff}), we see that
there are only two cases we need to consider.

These cases can be distinguished by two properties.  If
\begin{itemize}
\item   $f_n(-\Lambda R,u,j)$ remains finite as  $\Lambda R \to \infty$, and  
\item   $f_n(r,u,j)- f_0(r,u)$ remains finite as $\Lambda R\to \infty$
          for each value of $r$,
\end{itemize}
then our first case will apply.
Our second case will arise in all other situations, when either
one or both of these conditions fail.

In the first case,
$f_n(-\Lambda R,u,j)$ remains finite as  $\Lambda R \to \infty$.  
This situation arises when the UV divergence
from the four-momentum integration within $f_n$ is cancelled by the
diverging Kaluza-Klein number $r\equiv -\Lambda R\to -\infty$ 
in the denominator of the integrand.
In such cases, the second line of Eq.~(\ref{diff}) is 
finite by itself and may be identified as $\Delta_n$:
\beq
              \Delta_n ~=~ - \lim_{\Lambda R\to \infty}\,
              \frac{1}{|n|}\sum_{j = 0}^{|n| - 1} \int_{0}^{1}du \,
              f_n (-\Lambda R,u,j)~.
\label{Delt}
\eeq
The first line of Eq.~(\ref{diff}) must then also be
individually finite, which implies that 
the KK summation over the difference $f_n - f_0$ should also be finite
as $\Lambda R\to\infty$. 
By itself, this need not imply that each $f_n-f_0$ should be finite
for each individual term in the KK sum, for it is possible that
divergences of individual $f_n(r)-f_0(r)$ as $\Lambda R\to\infty$
are cancelled across the increasingly many terms in the sum as $\Lambda R\to\infty$.
(We shall see an explicit example of this phenomenon in Sect.~3.)
However, if we additionally know that 
each $f_n(r,u,j)- f_0(r,u)$ remains finite as $\Lambda R\to \infty$
          for each value of $r$ (our second defining criterion above),
it then follows that all regulator dependence must cancel
within the difference $f_n-f_0$ in Eq.~(\ref{diff}).  
In such cases, we can therefore proceed to identify $\alpha_0$ and $\alpha_n$ 
as the cutoff-independent parts of $f_0$ and $f_n$,
respectively. 

Alternatively, it may happen that one or both of the two conditions 
itemized above are not satisfied.
This is therefore the second possible case we need to face.
For example, if  $f_n(-\Lambda R,u,j)$ diverges as  $\Lambda R \to \infty$,
then neither expression within 
Eq.~(\ref{diff}) is finite by itself, and a further rearrangement of
terms within Eq.~(\ref{diff}) is needed.
However, we can generally handle this situation as follows.
In general, we can 
identify $\alpha_n$ as the  cutoff-independent part of the difference
$f_n - \tilde{f}_n$, where $\tilde{f}_n$ is the value of $f_n$
when all of the bare masses in our theory vanish and renormalization
conditions for massless particles have been applied.  Subtracting
$\tilde{f}_n$ from $f_n$ then cancels the cutoff-dependent terms  that do
not contain a bare mass, and subtracting $L_0$ from $L_n$ then cancels
whatever cutoff dependence remains. 
However, the price we pay is that these extra $\tilde f_n$ and $\tilde f_0$
terms are now shifted into $\Delta_n$, so that $\Delta_n$ is
now given by 
\beq
   \Delta_n ~=~  \lim_{\Lambda R \to \infty} 
              \frac{1}{|n|}\sum_{j = 0}^{|n| - 1}\int_{0}^{1}du 
    \left\lbrace
      \sum_{r = -\Lambda R}^{\Lambda R}
    \left[ \tilde{f}_n (r,u,j) - \tilde{f}_0(r,u) \right] 
        - f_n (-\Lambda R,u,j)
               \right\rbrace~.
\label{Delt2}
\eeq
Of course, these extra terms are precisely what are needed in order to
cancel the divergence of $f_n(-\Lambda R)$ as $\Lambda\to\infty$ in cases
in which it diverges, and render $\Delta_n$ finite.
Even when $f_n(-\Lambda R)$ remains finite as $\Lambda\to\infty$, 
these extra terms will preserve the finiteness of $\Delta$ and compensate
for the shifted definition of $\alpha$-functions relative to our first case above.

Even though $\Delta_n$ is finite in each case,
it is still important to be able to write $\Delta_n$ in an explicitly 
regulator-independent form.  We shall show how to do this in Sect.~3.

Thus, we conclude that as our regulators are removed, the difference
between loop diagrams will always take the form of Eq.~(\ref{nocut}) regardless
of whether $f_n (-\Lambda R,u,j)$ or $f_n(r,u,j)-f_0(r,u)$ have finite limits as 
$\Lambda R\to\infty$.  In Eq.~(\ref{nocut}),
all dependence on a cutoff has been absorbed into observed
parameters, and it is understood that the Kaluza-Klein $r$-summation in Eq.~(\ref{nocut})
is to be evaluated symmetrically, with equal and opposite diverging limits. 
Expressions
of this form  will then enable us to obtain regulator-independent
equations for  such renormalized KK parameters as masses and couplings.

It may seem suspicious that we have defined the $\alpha$- and $\Delta$-functions differently 
for the cases in which $f_n(-\Lambda R)$ 
and $f_n-f_0$ either remain finite or diverge as $\Lambda\to\infty$.
However, this was done simply as a matter of convenience.
In all cases, the most general procedure is the second one 
which we have outlined above,
and this procedure is also valid even when 
both of our defining criteria are met.
In such cases, this procedure merely introduces extraneous terms to the
$\alpha$- and $\Delta$-functions,  but these new additions
will always cancel in the loop diagram difference $L_n - L_0$.

Finally, before concluding, we remark that the procedure we have outlined
here has relied rather fundamentally on the assumption that the one-loop diagrams
we are regulating can be evaluated through the introduction of a single
Feynman parameter $x$ (or $u$).
However, this procedure readily generalizes to one-loop diagrams that would utilize
arbitrary numbers of Feynman parameters.
For example, in the case of two Feynman parameters,
we have already shown in Ref.~\cite{paper1a} that
the identity in Eq.~(\ref{kgenid}) generalizes to take the form
\beqn
     L_{n_1,n_2}(p_1,p_2)  ~=~
   \frac{i}{|n_1 n_2|} \sum_{j_1 = 0}^{|n_1| - 1}\, \sum_{j_2 = 0}^{|n_2| - 1}
            \int_{0}^{1}d{u}_1
            \int_{0}^{1}d{u}_2
     \sum_{{r} = -\Lambda R }^{\Lambda R}
          f_{n_i} (p_i, r, u_i, j_i) ~+~ E_{s_1,s_2}\nonumber\\
\label{Lformgen}
\eeqn
where  $s_i\equiv {\rm sign}(n_i)$ 
and $ u_i \equiv x_i |n_i| - j_i$.
We have also defined
$ \hat r\equiv r - s_1 j_1 - s_2 j_2$, and then dropped
the hat on $\hat r$.
The quantity $E_{s_1,s_2}$ in Eq.~(\ref{Lformgen}) represents a so-called ``endpoint''
contribution
[analogous to the final term in Eq.~(\ref{diff})] 
which depends on $f$ evaluated at or near the limits of the KK summation~\cite{paper1a}.
In such cases,
the $\alpha$-functions are defined analogously to the case of a 
single Feynman parameter, with the endpoint contributions $E$ leading to corresponding
$\Delta$-functions.
Indeed, the only changes to the basic formalism we have sketched are that
there are now two variables of integration, two mode-number indices for $f$, $\alpha$, and $\Delta$, 
and slightly less trivial endpoint contributions.

\section{Two Explicit Examples
\label{examples}}
\setcounter{footnote}{0}

In this section, we shall provide two explicit examples of 
the general procedure outlined in Sect.~2.
These two examples are designed to illustrate the two different cases
sketched at the end of Sect.~2.

\subsection{First example:  Pure scalar theory}

Our first example will assume that our external particles in Fig.~\ref{fig1}
are Lorentz scalars, and that the solid lines in Fig.~\ref{fig1} represent
scalars as well.
In this case, the generic diagram $L_n(p)$ is given by
\beq
  L_n (p) ~=~ \sum_r \int \frac{d^4 k}{(2\pi)^4} ~ \frac{1}{k^2 - r^2
          /R^2 - M^2} ~\frac{1}{(k - p)^2 - (r - n)^2 /R^2 - M^2}~
\label{gral}
\eeq
where $k$ is the four-momentum of a particle in our loop and $r$ is
its mode number.
Combining the denominators using standard Feynman-parameter methods,
we can then cast this expression into the form in Eq.~(\ref{kgenform})
where 
\beq
   f_n(p,r,x) ~\equiv~ \int \frac{d^4 \ell_E}{(2\pi)^4} 
     \left[\frac{1}{\ell_{E}^2 + {\ell^4}^2 + \mathcal{M}^2 (x)}\right]^2 ~,~~~~~~n\in\IZ~,
\label{intwick}
\eeq
where $\ell\equiv k-xp$ is the shifted
 {\it five}\/-momentum 
[\ie, $\ell^\mu\equiv k^\mu -xp^\mu$ and $\ell^4\equiv (r-xn)/R)$],
where $\ell_E$ is the standard Euclidean (Wick-rotated) version of $\ell$,
and where the effective mass in Eq.~(\ref{intwick}) is given by
\beq
         \mathcal{M}^2 (x) ~\equiv~ M^2 ~+~ x(x - 1)\left\lbrack p^2 -
         {n^2\over R^2} \right\rbrack~.
\label{delt}
\eeq

As it stands, these expressions are divergent.  We can regulate them,
while preserving the full {\it five}\/-dimensional Lorentz invariance,
using either of the two regulators introduced in Ref.~\cite{paper1a}.
Either regulator leads to an expression of the form in Eq.~(\ref{kgenform2}),
and after the variable substitutions in Eq.~(\ref{kgenid}),
these $f$-functions take the forms
\beqn
   f_0(r,u) &=& \int {d^4\ell_E\over (2\pi)^4}  \left\lbrack {1\over
                  \ell_E^2 + r^2/R^2+ M^2 + u(u-1)p^2 }\right\rbrack^2
                  \nonumber\\ 
   f_n(r,u,j) &=& \int {d^4\ell_E\over
                  (2\pi)^4}  \left\lbrack {1\over \ell_E^2 +
                  (r-u)^2/R^2+ M^2 + (u+j)(u+j-|n|) \left({p^2\over
                  n^2} - {1\over R^2}\right)}\right\rbrack^2.\nonumber\\
\label{fexamples}
\eeqn
In the case of the EHC regulator, the domain of integration
for $f_0$ and $f_n$ 
within Eq.~(\ref{intwick}) is regulated according to the prescriptions~\cite{paper1a}
\beqn
   f_0:~~~~~~  \ell_E^2 & \leq&  \Lambda^2 - r^2/R^2\nonumber\\
   f_n:~~~~~~  \ell_E^2 & \leq& \Lambda^2 - (r-u)^2/R^2~,
\label{kdomains}
\eeqn
while in the case of the EDR regulator, we merely shift the measure
$d^4 \ell_E / (2\pi)^4 \to d^{4-\epsilon} \ell_E/(2\pi)^{4-\epsilon}$ where
$\epsilon$ and $\Lambda$ are related according to~\cite{paper1a}
\beq
     {2\over\epsilon} - \gamma + \log(4\pi) +
       \calO(\epsilon) ~=~ 2\,\log(\Lambda R) + \delta~.
\label{eps_lam}
\eeq
Here $\gamma$ is the Euler-Mascheroni constant,
and $\delta$ is an inconsequential parameter which vanishes as 
$\Lambda\to \infty$ (or as $\epsilon\to 0$).
In either case, the prescriptions in Eq.~(\ref{kdomains}) or
(\ref{eps_lam}) are precisely designed to preserve five-dimensional
Lorentz invariance~\cite{paper1a}.  

Using either regulator, these functions $f_0$ and $f_n$ can be evaluated
explicitly.  
Applying appropriate renormalization conditions in each case
(for example, $p^2=M_e^2$ and $p^2= M_e^2 + n^2/R^2$ respectively,
where $M_e$ is the bare four-dimensional mass of the external particles
in Fig.~\ref{fig1})
and explicitly performing the integrals in
Eq.~(\ref{fexamples}) with the EHC regulator, we obtain
\beq
  f_0(r,u)~=~ {1\over 16\pi^2}\, \left\lbrace
            {r^2-\Lambda^2 R^2\over [\Lambda^2 + \calM^2(u)] R^2} +
         \log[(\Lambda^2+\calM^2(u))R^2] - \log[r^2 + \calM^2(u) R^2] \right\rbrace
\label{kintermed}
\eeq
where $\calM^2(u)\equiv M^2 + u(u-1)M_e^2$.
The function $f_n(r,u,j)$ is given by an identical expression
with the replacements $r\to\rho\equiv r-u$ and $u\to y\equiv (u+j)/|n|$. 
As promised, we see that the $\Lambda$-dependence is completely cancelled
in the difference $f_n-f_0$ as $\Lambda\to \infty$.

Given the forms in Eq.~(\ref{kintermed}), it may easily be verified
that $f_n(-\Lambda R,u,j)$ remains finite (and in fact vanishes) as $\Lambda R\to\infty$.
Likewise, it is easy to check that
$f_n(r,u,j)-f_0(r,u)$ remains finite as
$\Lambda R\to\infty$ for each value of $r$.
This is therefore an example of the first case discussed at the end
of Sect.~2, whereupon we see that we can
identify $\alpha_0$ and $\alpha_n$ as the cutoff-independent
parts of $f_0$ and $f_n$, respectively.
Looking at Eq.~(\ref{kintermed}), we see that 
the only cutoff-independent term that survives 
in the difference $f_n-f_0$ as $\Lambda\to \infty$
is the final term in Eq.~(\ref{kintermed}).
We can therefore identify
\beqn
     \alpha_0(r,u)&=& -{1\over 16\pi^2} \, \log\left[ r^2    + \calM^2(u) R^2 \right]\nonumber\\
   \alpha_n(r,u,j)&=& -{1\over 16\pi^2} \, \log\left[ \rho^2 + \calM^2(y) R^2 \right]~
\eeqn
where $\rho\equiv r-u$ and $y\equiv (u+j)/|n|$.

We can also explicitly evaluate $\Delta_n$ for this example. 
Since $f_n(-\Lambda R,u,j)$ remains finite as $\Lambda R\to\infty$,
we know that $\Delta_n$ is given by Eq.~(\ref{Delt}). 
However, since $f_n(-\Lambda R,u,j)$ actually vanishes as $\Lambda R\to\infty$,
Eq.~(\ref{Delt}) implies that $\Delta_n$ vanishes as well.
We therefore conclude that $\Delta_n=0$ for this example.
Indeed, we have found this to be a common result for theories without gauge invariance
(for which our EHC regulator is appropriate).
However, these results are of course independent of the specific regulator employed
as long as the regulator respects all of the five-dimensional symmetries that exist 
in the theory prior to compactification.

\subsection{Second example:  Five-dimensional QED}

As a somewhat more complicated example, let us 
now consider the case of a vacuum polarization diagram in five-dimensional QED.
We can therefore take 
the external lines in Fig.~\ref{fig1} to correspond to an incoming/outgoing KK photon, 
while our internal lines correspond to a tower of KK fermions.
As in the previous example, we shall assume that this tower of KK fermions has
bare five-dimensional mass $M$.
However, unlike the situation in the previous example, this 
theory has both five-dimensional Lorentz invariance
and five-dimensional gauge invariance prior to compactification.

Because the incoming and outgoing photons carry five-dimensional Lorentz
vector indices $M,N=0,1,2,3,4$, this diagram $L_n^{MN}$ will have a 
Lorentz two-tensor structure. 
Thus, in general, after integration over the internal momentum 
running in the loop, $L_n^{M N}$ will contain
one part which is proportional to $g^{M N}$ and another which is
proportional to $p^M p^N$. 
We shall restrict our attention to the part of
$L_n^{\mu \nu}$ which is proportional to $g^{\mu \nu}$, since this is the component
which gives rise to renormalizations of the squared masses of the KK photons. 
We shall henceforth denote these terms as $\tilde L_n^{\mu\nu}$.

Clearly, $\tilde{L}_0^{\mu \nu}$ vanishes 
for the photon zero mode;  this is because 
four-dimensional gauge invariance protects the zero-mode photon 
from gaining a mass. 
However, using the techniques we have sketched above, it is relatively
straightforward to calculate $\tilde L_n^{\mu\nu}$ for non-zero $n$. 
Employing the EDR regularization procedure from Ref.~\cite{paper1a}
(as appropriate for theories with gauge invariance)
and applying the renormalization condition $p^2-n^2/R^2=0$ for on-shell
external KK photons (as appropriate for calculating mass corrections),
we then find that $\tilde L_n^{\mu\nu}$ takes the form
in Eq.~(\ref{kgenform2}) with
\beq
    f^{\mu\nu}_n(r,x)~\equiv~ -{e^2 R^\epsilon\over 4\pi^2 R^2}
      \, (2x-1) \, n \, (r-xn) \, W\, g^{\mu\nu}~,~~~~~~~~ n\not= 0~,
\label{kpa}
\eeq
where
\beq
     W ~\equiv~ \frac{2}{\epsilon} - \gamma + \log(4\pi) -
         \log[(r-xn)^2 + (MR)^2] +  \mathcal{O}(\epsilon)~.
\label{kdform3}
\eeq
Equivalently, after the variable substitutions in Eq.~(\ref{kgenid})
and dropping the hats on $\hat r$ and $\hat u$,
we find that $f_n^{\mu\nu}(r,u,j)$ is given by Eqs.~(\ref{kpa}) and
(\ref{kdform3})
where we simply replace $x\to y\equiv (u+j)/|n|$ and
$r-xn\to (r-u) \cdot {\rm sign}(n)$.

Given these results, 
we can now examine the behavior of $f_n^{\mu\nu}(-\Lambda R, u, j)$ 
as $\Lambda R\to\infty$ and $\epsilon\to 0$.
It first glance, it may appear that this expression diverges.
However, we must recall that $\Lambda R$ and $\epsilon$ are
related~\cite{paper1a} according to Eq.~(\ref{eps_lam}).  Substituting Eq.~(\ref{eps_lam})
into Eq.~(\ref{kdform3}),
we find that $W\approx -2u/(\Lambda R)+\delta$ as $\Lambda R\to\infty$.
Since $\delta$ also vanishes as $\Lambda R\to \infty$ [and generally does so more
quickly than $1/(\Lambda R)$],
we see that $f_n^{\mu\nu}(-\Lambda R, u, j)$ actually remains finite 
in this limit.
On the other hand, it is immediately apparent that $f_n(r,u,j)$ diverges
as $\Lambda R\to\infty$, while $f_0(r,u)=0$.
Thus, our second defining condition at the end of Sect.~2 is not satisfied,
whereupon we see that five-dimensional QED is an example of the second case discussed
at the end of Sect.~2. 

Before going further,
it is important to stress that these observations do not contradict
the overall finiteness of Eq.~(\ref{diff}).
Indeed, the finiteness of 
$f_n^{\mu\nu}(-\Lambda R, u, j)$ in this limit implies that the second
line of Eq.~(\ref{diff}) is finite by itself, and this
in turn implies that the first line of Eq.~(\ref{diff}) must also be
finite as $\Lambda R\to\infty$.
In order to see how these divergences cancel,
let us consider the simple case where $n=1$.
In this case, $j=0$ and $y=x$, whereupon
we find that the first line of Eq.~(\ref{diff}) takes the form
\beq
         \sum_{r= -\Lambda R}^{\Lambda R}  \, \int_0^1 du ~
        (2 u-1) (r-u) \left\lbrace
         \log [(\Lambda R)^2] - \log[ (r-u)^2 + (MR)^2 ] \right\rbrace~.
\label{kexamp}
\eeq
Note that both terms in Eq.~(\ref{kexamp}) diverge like
$(\Lambda R)\log(\Lambda R)$ as $\Lambda R\to \infty$;
this is a subtle interplay between terms in which the cutoff $\Lambda R$
appears explicitly in the integrand/summand (as in the first term above) 
and in which the cutoff appears only as the upper limit on the $r$-summation
(as in the second term above). 
It is nevertheless straightforward to verify that these two divergences
cancel directly in Eq.~(\ref{kexamp}), leading to a finite expression
as $\Lambda R\to\infty$.  A similar cancellation happens for each
value of $n$.
             
According to our general prescription in Sect.~2.1, we must therefore
identify $\alpha_n(r,u,j)$ as the cutoff-independent part of the
difference $f_n-\tilde f_n$, where $\tilde f_n$ is the value of $f_n$
when all of the bare masses in our theory vanish.
We thus find that
\beq
    \alpha^{\mu\nu}_n(r,u,j)~ =~ \frac{e^2 g^{\mu \nu}}{4\pi^2 R^2}\,
            (2y-1)\, |n|\, (r-u)\, \left\lbrace \log \left[ (r - u)^2 + (M R)^2 \right]
                 - \log \left[ (r - u)^2 \right] \right\rbrace
\label{alpha_QED}
\eeq
for all $n\not=0$, while $\alpha_0^{\mu\nu}=0$.
We can also calculate $\Delta_n^{\mu\nu}$.
Using the definition in Eq.~(\ref{Delt2}) and incorporating
the relation in Eq.~(\ref{eps_lam}), we find
\beqn
\Delta^{\mu\nu}_n &=& - {e^2 g^{\mu\nu}\over 4\pi^2 R^2}\,
       {1\over |n|} \,\sum_{j=0}^{|n|-1} \,\int_0^1 du\,\nonumber\\
     && ~~~ 
       \lim_{\Lambda R\to \infty}\,
      \sum_{r= -\Lambda R+1}^{\Lambda R}\,
       (2y-1)\,|n|\, (r-u) \left\lbrace
         \log[ (\Lambda R)^2 ] 
       - \log[ (r-u)^2 ] \right\rbrace~.~~~~~~
\label{D_QED}
\eeqn

While these are indeed the correct results, 
we note that the summand/integrand in 
Eq.~(\ref{D_QED}) is still regulator-dependent,
depending explicitly on $\Lambda$. 
Thus, in contrast to the regulator-independent 
$\alpha^{\mu\nu}$-terms from Eq.~(\ref{alpha_QED}), 
we see that we have not yet succeeded in writing 
$\Delta^{\mu\nu}_n$ in a manifestly 
finite, {\it regulator-independent}\/ manner. 
We emphasize that it is not the cutoffs on the upper and lower limits 
of the KK sum in Eq.~(\ref{D_QED}) which cause difficulty;
after all, at an algebraic level, these KK summation cutoffs may 
be smoothly removed without difficulty.  
By contrast, it is the explicit factor of $\Lambda R$
within the summand itself 
which causes algebraic difficulty and which prevents this expression
from being truly regulator-independent.

Our goal, of course, is to show that
the expression for $\Delta^{\mu\nu}_n$ in 
Eq.~(\ref{D_QED}) --- just like our expressions for 
$\alpha_n^{\mu\nu}$ and $\alpha_0^{\mu\nu}$ ---
can be rewritten in a manifestly finite, regulator-independent manner. 
In other words, without affecting the value of $\Delta^{\mu\nu}_n$,
we wish to replace the second line of Eq.~(\ref{D_QED})
with an expression of the general form
\beq
      \sum_{r= -\infty}^\infty\,  h(r,u,j)
\label{hdef}
\eeq
where $h(r,u,j)$ is a regulator-independent function.

Clearly, in order to derive the appropriate function $h(r,u,j)$,
we need to find a way of algebraically redistributing the explicit 
$\Lambda$-dependence in the summand of Eq.~(\ref{D_QED})   
across all of the terms in the KK sum.
This can be accomplished as follows.
First, we observe that explicitly performing the KK summation 
and Feynman integration for the $\Lambda$-dependent term in Eq.~(\ref{D_QED}) yields 
\beq
     \int_{0}^{1}du\, 
  \sum_{r = -\Lambda R + 1}^{\Lambda R} 
    (2y - 1)|n|(r - u) \log[(\Lambda R)^2] ~=~ -\frac{1}{3} (\Lambda R)\,
             \log[(\Lambda R)^2]~.
\label{mu_nu_sum}
\eeq
Note that this result is an odd function of the cutoff $\Lambda R$.
However, we can now ``invert'' this and rewrite 
any odd function of the cutoff $F(\Lambda R)$
in the desired form using the identity
\beqn
      F(\Lambda R) &=& \phantom{-}\half \, 
          \left[ F(\Lambda R) - F(-\Lambda R) \right] \nonumber\\
  &=&  \phantom{-}\half \, \int_{-\Lambda R}^{\Lambda R} dz\, f(z)~~~~~~~~~~~
               ~~~~~~~~~ {\rm where}~~ f(z)\equiv dF(z)/dz\nonumber\\
  &=&  \phantom{-}\half \, \sum_{r= -\Lambda R+1}^{\Lambda R} \, 
             \int_{r-1}^r dz\, f(z)\nonumber\\
  &=&  -\half \, \sum_{r= -\Lambda R+1}^{\Lambda R} \, \int_1^0 du \, f(r-u)
                         ~~~~~~ {\rm where}~~ u\equiv r-z\nonumber\\
  &=& \phantom{-}\half \, \sum_{r= -\Lambda R+1}^{\Lambda R} \, \int_0^1 du\, f(r-u)~.
\eeqn
As we see, this identity has the net effect of 
throwing an arbitrary, explicit (odd) dependence
on $\Lambda R$ into the upper and lower limits of a discrete sum,
just as desired. 
In the case at hand, we have $F(z)\equiv -\third z \log(z^2)$, whereupon
we see that $f(z)\equiv  -\third (\log z^2 +2)$.
We thus find that $h(r,u,j)\equiv -\third (\log[(r-u)^2] + 2)$ in 
Eq.~(\ref{hdef}), whereupon we conclude that
$\Delta^{\mu\nu}_n$ can be written in the 
regulator-independent form
\beq
\Delta^{\mu\nu}_n ~=~  {e^2 g^{\mu\nu}\over 4\pi^2 R^2}\,
       {1\over |n|} \,\sum_{j=0}^{|n|-1} \,\int_0^1 du\,
      \sum_{r= -\infty}^{\infty}\,
           \left\lbrace
     \left[ (2y-1)\,|n|\, (r-u) + {1\over 6} \right] \log[ (r-u)^2 ]  
           + {1\over 3} \right\rbrace~.
\label{D_QED2}
\eeq
Similar algebraic manipulations can also be performed for other
diagrams of interest.

It is straightforward to verify that the expression in Eq.~(\ref{D_QED2}) 
converges to a finite result.  As an example, let us consider
the case with $n=1$.  For $n=1$, we have $j=0$ and $y=u$.  Explicitly performing
the $u$-integration and defining $s\equiv r-1/2$, we then find
\beq
   \Delta^{\mu\nu}_1 ~=~ 
       {e^2 g^{\mu\nu}\over 144\pi^2 R^2}\, \sum_{s\in \IZ+1/2}\,
         \left\lbrace
          (4-24 s^2) + 3s(4s^2-1) \left[
             \log[(s+\half)^2] - \log[(s-\half)^2] \right]
           \right\rbrace~.
\label{sform}
\eeq
We immediately observe that the summand has a symmetry under $s\to -s$,
which implies that contributions from positive values of $s$
are identical to contributions from negative values of $s$.
Thus, the finiteness of Eq.~(\ref{sform}) does {\it not}\/ rely on a cancellation
between contributions from positive and negative KK mode numbers;  sums
over positive or negative values of $s$ are each {\it separately}\/ convergent.  
Furthermore, we see that the contributions from $|s|=1/2$ are also finite, since
the divergent logarithm in Eq.~(\ref{sform}) for $|s|=1/2$ is multiplied by the 
factor $(4s^2-1)$, which vanishes even more strongly.  
This cancellation of the logarithmic divergence can also be 
verified by evaluating the original integral in Eq.~(\ref{D_QED2})
directly with $r=0,1$. 
Finally, for large $|s|$, it is straightforward
to verify that the summand in Eq.~(\ref{sform}) scales as $\sim 1/s^2$.
Thus the KK sum in Eq.~(\ref{sform}) is absolutely convergent, as required. 
In fact, it can easily be shown that the $s$-summation in Eq.~(\ref{sform})
converges to $-36 \zeta(3)/\pi^2$, where
$\zeta(n)$ is the Riemann zeta-function.
We therefore find $\Delta_1^{\mu\nu} = - [e^2 g^{\mu\nu}/(4\pi^4 R^2)] \zeta(3)$,
in agreement with results quoted in Refs.~\cite{paper1a,CMS}.
It turns out that $\Delta_n^{\mu\nu}$ takes this value for
all $n\not=0$.

\section{Effective Field Theories for Kaluza-Klein Modes 
\label{EFT}}
\setcounter{footnote}{0}

Thus far, we have described how to calculate radiative shifts to 
physical KK parameters such as KK masses and couplings.
In doing this, we have included effects from all 
energy scales from the deep infrared to the ultraviolet, as required.

However,
as a question of both practical importance and mathematical
curiosity, 
it is useful to have an effective field theory (EFT) description
of our KK system which is appropriate for any arbitrary finite
cutoff $\mu$.
In the context of KK theories,
EFT's are particularly useful tools for doing calculations
and making predictions because they 
only contain finite numbers of KK states, and the presence
of a cutoff eliminates problems of non-renormalizability.
Indeed, it is ultimately only an EFT (with finitely many relevant parameters)
which can be fit to experiment.  

Towards this end, 
we now seek to obtain EFT's which can be used to describe our KK systems
at lower energy scales.
In order to do so,
we need to determine how these radiative corrections accrue as we move from
the ultraviolet limit (where the full 5D Lorentz invariance of the
theory is restored and the corrections vanish) to the infrared. 
In other words, we seek to express these radiative corrections as
evolution functions of the energy scale $\mu$ at which a collider
might operate.  

In this section, we will present a procedure for doing this which is
based on a Wilsonian approach.
We begin, in Sect.~4.1, with some general comments concerning 
novel features of the Wilsonian approach which arise for KK theories. 
In Sect.~4.2, we then present our general results describing the 
flow of KK parameters using a Wilsonian treatment.

\subsection{Wilsonian flow in Kaluza-Klein theories}

Given a Lagrangian in the UV limit,
a Wilsonian 
approach to analyzing the behavior of the corresponding
effective field theories at
different lower energy scales $\mu$ consists of
integrating out all physics
with (Euclidean) momentum or energy scales exceeding $\mu$.
In general, this has two effects:  it produces new effective interactions
which were not present in the original UV Lagrangian, and
it changes the  values of the bare parameters which were already
present in the original
Lagrangian, rendering them $\mu$-dependent.
The resulting Lagrangian then describes a Wilsonian EFT
at energy scale $\mu$.
Within the framework of such a theory, we would then perform calculations
in the usual way based on this effective Lagrangian,
except that $\mu$ now serves as a hard UV cutoff for
such calculations.
This is appropriate because the contributions from the physics 
at scales above $\mu$ has already been absorbed directly into the 
EFT Lagrangian.

In an ordinary four-dimensional setup,
this process of integrating out physics above the scale $\mu$ 
is performed uniformly across all sectors of the theory, for all
particle species that may appear in the Lagrangian.
Following the strict approach outlined above, 
we do not eliminate heavy particles from our theory;  it is
simply that their contributions become small because of the kinematic
constraints that operate within a restrictive cutoff $\mu$.
For example, let us imagine that our 4D theory contains different particle 
species with masses $m_i$ for $i=1,..,n$. 
Evaluating a one-loop radiative correction 
within such a theory, we would sum over the
contributions from each particle which is allowed to propagate in the loop.
Likewise, each of these contributions is evaluated by integrating the possible
loop momentum over all possible values up to infinity.
Of course, for each individual particle running in the loop,
the contributions to the momentum integral will be 
greatest from those values of the loop momentum for which
the internal particle is closest to being on-shell.
As a result, the contributions from particles whose masses exceed the cutoff
will be exceedingly 
small because they will be significantly off-shell for all allowed values  
of the loop momentum below the cutoff. 
However, within the framework of a Wilsonian-derived EFT, 
we are only instructed to truncate each of these momentum integrals   
at a scale $\mu$.  
We are {\it not}\/ instructed to eliminate any of the particle 
species themselves, even if their masses $m_i$ significantly exceed 
$\mu$.  

This situation changes significantly for KK theories.
At first glance, one might think that a KK theory is simply another
four-dimensional theory with an infinite set of increasingly heavy particles.
However, we must remember that in a KK theory, the masses of these particles 
receive contributions from (and are therefore the reflections of)
the fifth components of a higher-dimensional
momentum.  It would be acceptable to disregard this fact if we were not
aiming to develop EFT's that respect our 
original higher-dimensional symmetries as far as possible.
This includes higher-dimensional Lorentz invariance.
However, in the present case, we seek to follow an intrinsically
higher-dimensional approach so as to avoid the introduction of
spurious Lorentz-violating contributions.

As a result, we must follow an intrinsically {\it higher-dimensional}\/ Wilsonian
approach to deriving our EFT's.
In the case of five dimensions,
this means that one-loop integrals such as that in 
Eq.~(\ref{intwick}) must be truncated with an intrinsically
 {\it five-dimensional}\/ cutoff $\mu$.  In other words,
we must impose the five-dimensional Lorentz-invariant constraint 
\beq
            \ell_E^2+ (\ell^4)^2 ~\leq~ \mu^2~.
\label{5Dmucutoff}
\eeq
Because $R \ell^4\equiv r-xn$, 
such a constraint equation
correlates the cutoff for the integration  over the
four-momentum $\ell_E$ with the cutoff for the summation over the KK
index $r$.  In particular, the constraint in Eq.~(\ref{5Dmucutoff}) can be
implemented by restricting the KK summation to integers in the range
\beq
           -\mu R + xn ~\leq~ r ~\leq~ \mu R + xn~
\label{murcut}
\eeq
and then restricting our $\ell_E$-integration to the  corresponding
range
\beqn
            \ell_{E}^2 ~& \leq&~  \mu^2 - ({\ell^4})^2 \nonumber\\ &
                        \leq&~  \mu^2 - (r-xn)^2/R^2~.
\label{mukcut}
\eeqn
Thus, we see that in a truly five-dimensional setup, 
a Wilsonian treatment not only implies a truncation for 
four-dimensional loop integrations;
 {\it it also implies a truncation in the KK tower}\/.
In other words, we not only eliminate certain momentum scales
from consideration;  we also eliminate heavy KK states entirely.

It is, of course, no accident that the KK constraint in Eq.~(\ref{murcut})
resembles that in Eq.~(\ref{kgenform2}),
and that the four-momentum cutoff in Eq.~(\ref{mukcut}) resembles that 
in Eq.~(\ref{kdomains}), except with the replacement $\Lambda\to\mu$.
Eqs.~(\ref{kgenform2}) and (\ref{kdomains}) together stem
from our extended hard-cutoff (EHC) regulator, whose defining
characteristic is also the imposition of a five-dimensional Lorentz-invariant
cutoff. 
There is, however, a major conceptual difference between the two cutoffs 
$\Lambda$ and $\mu$:
while $\Lambda$ is a {\it regulator}\/ cutoff 
which is always removed at the end of a calculation, 
$\mu$ is a finite {\it physical}\/ (Wilsonian) cutoff 
which defines an associated momentum scale for our EFT
and which serves as a finite physical cutoff for calculations
performed within the context of that EFT.  Such a cutoff is not taken
to infinity at the end of such an EFT calculation.
It is therefore only at the algebraic level that these two cutoffs appear similar.

\subsection{Deriving Wilsonian EFT's}

Given these observations,
it is now straightforward to derive our Wilsonian EFT's 
as a function of the momentum scale $\mu$.
We shall do this to one-loop order, and 
shall provide a general method of deriving the $\mu$-dependence 
that our KK parameters (masses and couplings) accrue.
There are, of course, 
new effective operators that will also
be generated in these EFT's.  
The procedure we shall be outlining below also applies 
to their coefficients as well.\footnote{
   Of course, even our original higher-dimensional theory can be viewed as an EFT
   of its own, and hence would contain all possible operators
   consistent with our UV symmetries.  From this perspective,
   the only ``new'' operators that would be generated are those
   which break the higher-dimensional symmetries but not the
   four-dimensional symmetries.  In any case, the procedure
   we are outlining here is applicable to the coefficients
   of any operators, whether higher-order or not.
   However, for convenience, we shall refer
   to these coefficients as KK masses and couplings in what follows.}

We shall let $\lambda_n$ collectively denote these KK parameters.
Of course, these parameters $\lambda_n$ will receive
classical rescalings in cases where they are dimensionful.
In order to eliminate these classical rescalings, we shall henceforth
assume that each $\lambda_n$ has been multiplied by a sufficient
power of the radius $R$ so as to be dimensionless.
Likewise, we know that each $\lambda_n$ will receive quantum
corrections which are divergent when our cutoffs are removed;
indeed, it is only differences such as $\lambda_n-\lambda_0$ 
which can be expected to remain finite.
Our strategy will therefore be to derive the $\mu$-dependence
of differences such as $\lambda_n-\lambda_0$.
The parameters $\lambda_{n\not=0}$ describing the excited KK states in our 
EFT associated with any scale $\mu$ can then be obtained
in terms of the parameters $\lambda_0$ for the zero modes.

Let us first consider the full, one-loop radiative correction 
to the difference $\lambda_n-\lambda_0$.
According to Eq.~(\ref{nocut}), this one-loop radiative correction
$\Delta(\lambda_n-\lambda_0)$ is given by
\beq
  \Delta(\lambda_{n} - \lambda_{0}) ~=~ 
    \sum_{ r = -\infty}^\infty \frac{1}{|n|}\sum_{j = 0}^{|n| -1}
              \int_{0}^{1}d u \, 
        \left\lbrack \alpha_n (r, u,j) - \alpha_0 (r, u) \right\rbrack
      + \Delta_n
\label{infraredlimit}
\eeq
where the $\alpha_n$- and $\Delta_n$-functions are determined
according to the procedures described in Sect.~2.
Note that since the $\lambda_n$ are dimensionless, the $\alpha_n$-functions
corresponding to these $\lambda_n$ 
are dimensionless as well.
Clearly Eq.~(\ref{infraredlimit}) represents the complete
radiative shift in $\lambda_n -\lambda_0$ that is accrued from momentum
scales all the way from the ultraviolet limit to the infrared.

Our goal, however, is to evaluate the corrections to the bare
parameters in the UV Lagrangian that accrue due to integrating
out only that portion of the physics associated with momenta exceeding $\mu$.
We therefore wish to calculate the {\it partial}\/ radiative correction 
from the ultraviolet limit
down to an arbitrary non-zero momentum scale $\mu$. 
To do this, we calculate the same one-loop radiative correction,
only now imposing the constraint
\beq
            \ell_E^2+ (\ell^4)^2 ~\geq~ \mu^2~,
\label{IRcutoff}
\eeq
where $\mu$ is treated as an {\it infrared}\/ cutoff.
In other words, we must integrate out the contributions from KK modes
satisfying
\beq
        r~\leq~ -\mu R + xn ~~~~~{\rm or}~~~~~
        r~\geq~  \mu R + xn~,
\label{IRcutoff2}
\eeq
where the corresponding $\ell_E$-integrations are restricted to the region
\beq
            \ell_{E}^2 ~ \geq~  \mu^2 - (r-xn)^2/R^2~.
\label{IRcutoff3}
\eeq
In analogy with Eqs.~(\ref{prediff}) and (\ref{diff}), 
we see that this then leads to the partial radiative correction
\beqn
  \Delta(\lambda_{n} - \lambda_{0})\bigg|_\mu   &=&
         \frac{1}{|n|}\sum_{j = 0}^{|n| - 1}\int_{0}^{1}du\, 
       \lim_{\Lambda R\to\infty}\,
       \Biggl\lbrack
          \left( \sum_{r= -\Lambda R+1}^{-\mu R} + \sum_{r= \mu R+1}^{\Lambda R}\right)
              f_n (r;\mu,\Lambda)\nonumber\\ 
        &&~~~~~~~~~~~~~~~-~
          \left( \sum_{r= -\Lambda R}^{-\mu R} + \sum_{r= \mu R}^{\Lambda R}\right)
              f_0 (r;\mu,\Lambda)  
               ~+~ \delta_{\mu R,0} \,f_0(0;\mu,\Lambda) 
       \Biggr\rbrack~\nonumber\\
 &=&  \frac{1}{|n|}\sum_{j = 0}^{|n| - 1}\int_{0}^{1}du\,
       \lim_{\Lambda R\to\infty}\,
       \Biggl\lbrace
          \left( \sum_{r= -\Lambda R}^{-\mu R} + \sum_{r= \mu R}^{\Lambda R}\right)
           \biggl\lbrack f_n (r;\mu,\Lambda) - f_0 (r;\mu,\Lambda)  \biggr\rbrack \nonumber\\
        &&~~~~~~~~~~~~~~~-~
              f_n (-\Lambda R;\mu,\Lambda)  
            - f_n (\mu R;\mu,\Lambda)  
               + \delta_{\mu R,0} \,f_0(0;\mu,\Lambda) 
                      \Biggr\rbrace~.\nonumber\\
\label{diff2}
\eeqn
Here the $f_n$-functions are the dimensionless functions appropriate 
for calculations of the $\lambda_n$ parameters.
Note that in writing these functions, we have suppressed their $(u,j)$ arguments
relative to the notation in previous sections.
However, we have also added two explicit
cutoff arguments, writing $f(r; \mu_1,\mu_2)$ where $\mu_1$ and $\mu_2$ respectively 
represent the infrared and ultraviolet cutoffs
which truncate the four-momentum integrals contained within these functions.
In Eq.~(\ref{diff2}), we have also defined $\delta_{\mu R,0}\equiv 1$ 
only if $\mu R=0$, and zero otherwise.
Terms proportional to $\delta_{\mu R,0}$ in Eq.~(\ref{diff2}) 
compensate for the notational overcounting which arises
in the $\mu R=0$ special case.  Finally, note that the $\mu R \to 0$ limit
of Eq.~(\ref{diff2}) indeed reproduces the full radiative correction
in Eq.~(\ref{diff}), which may then be replaced by the explicitly finite
form in Eq.~(\ref{nocut}).

In the expressions in Eq.~(\ref{diff2}), the scale $\mu$ 
always appears as as {\it infrared}\/ cutoff for the $f$-functions.
However, there also exists an alternative but equivalent set of expressions
in which $\mu$ appears as an {\it ultraviolet}\/ cutoff within the $f$-functions.
This alternative formulation exists because
  $\Delta(\lambda_{n} - \lambda_{0})|_\mu$   
can equivalently be calculated
by integrating all the way down to zero energy,
yielding the full contribution in Eq.~(\ref{nocut}),
but then subtracting the contributions that emerge
if we restore the partial contributions from zero energy back up
to $\mu$.
These latter contributions are given in 
Eq.~(\ref{diff}), where ultraviolet cutoff $\Lambda$ is replaced by $\mu$
and kept finite.
Combining the results in Eqs.~(\ref{nocut}) and (\ref{diff}),
we therefore have
\beqn
  \Delta(\lambda_{n} - \lambda_{0})\bigg|_\mu  &=& 
           \phantom{-}~ \frac{1}{|n|}\sum_{j = 0}^{|n| -1} \int_{0}^{1}d u \, 
    \sum_{ r = -\infty}^\infty\, 
        \left\lbrack \alpha_n (r, u,j) - \alpha_0 (r, u) \right\rbrack
      ~+~ \Delta_n\nonumber\\
          && -~   
         \frac{1}{|n|}\sum_{j = 0}^{|n| - 1}\int_{0}^{1}du ~ \biggl\lbrace
          \, \sum_{r = -\mu R}^{\mu R}\,
         \biggl\lbrack f_n (r;0,\mu) -f_0(r;0,\mu)\biggr\rbrack\nonumber\\
          && ~~~~~~~~~~~~~~~~~~~~~~~~~~~~~~~~~~~~~-~  f_n (-\mu R;0,\mu)\biggr\rbrace~.
\label{diff3}
\eeqn
Note that $\mu$ now appears as an ultraviolet cutoff within the $f$-functions.
For example, in the case of the first example discussed in Sect.~3,
we see that $f_0(r;0,\mu)$ is given by the expression in
Eq.~(\ref{kintermed}) with the replacement $\Lambda\to\mu$.
Note that in the final line of Eq.~(\ref{diff3}), we do {\it not}\/
take the additional step of replacing our $f$-functions with 
$\alpha$-functions in which the $\mu$-dependence 
is eliminated.  While such a replacement would have been appropriate
in the $\mu\to\infty$ limit, we are regarding $\mu$ as an
arbitrary finite parameter in Eq.~(\ref{diff3}).
Consequently, although any divergent behavior as a function of $\mu$
indeed continues to cancel in the difference $f_n -f_0$,
there can be subleading terms which scale with inverse powers of
$\mu$.  Such terms must be retained for finite $\mu$.
Also note that since $f(r;0,0)=0$,  
the $\mu\to 0$ limit of Eq.~(\ref{diff3})
again restores the full radiative contribution in Eq.~(\ref{nocut}).

In either case, with 
  $\Delta(\lambda_{n} - \lambda_{0})|_\mu$   
written as in Eq.~(\ref{diff2}) or as in Eq.~(\ref{diff3}),
we can then write the value of $\lambda_n$
at scale $\mu$ in terms of $\lambda_0$ at the same scale: 
\beq
    \lambda_n |_{\mu} ~=~  \lambda_0 |_{\mu} + 
         \left[ \kappa_n + \Delta(\lambda_n - \lambda_0 )\biggl|_{\mu} \right] 
\label{ren_lam}
\eeq
where $\kappa_n$ denotes the difference $\lambda_n  - \lambda_0 $ evaluated in the
UV limit: 
\beq
   \kappa_n  ~\equiv ~ 
    \left( \lambda_n - \lambda_0 \right) \biggl|_{\mu = \infty}~.
\label{kap}
\eeq
For example, if $\lambda$ corresponds to the squared (dimensionless) 
masses $ (mR)^2$ of the KK modes arising from a one-dimensional compactification
on a circle, then $\kappa_n= n^2$.

We conclude this section with two related comments.
First, it is clear that the above analysis has been based on our EHC regulator,
as introduced in Ref.~\cite{paper1a}.  This does not imply, however, that such a formalism
cannot be used in theories with gauge invariance.
If the higher-dimensional theory in question has gauge invariance,
this formalism may continue to be utilized;  the only difference is that
gauge-dependent counterterms must be introduced in order to compensate
for the (apparent)  
breaking of gauge invariance induced by the presence of a hard cutoff.
Moreover, an analogous formalism can be developed
which avoids the hard cutoff altogether, and which utilizes our EDR regulator
from Ref.~\cite{paper1a}.  This formalism would preserve our higher-dimensional
symmetries, both Lorentz invariance and 
gauge invariance, in a completely manifest way. 
Note that regardless of which specific regulator from Ref.~\cite{paper1a} is employed,
the crucial ingredient that enables this formalism to operate
is the fact that such regulators preserve the original symmetries of
our {\it higher-dimensional}\/ Lagrangian, and not merely those four-dimensional
symmetries which remain after compactification. 

Second, we also emphasize that the Wilsonian approach we have followed
here is different from the more canonical approach in which the renormalization
scale $\mu$ is introduced through renormalization conditions involving
the momenta of external particles.
By contrast, the above approach does not impose any particular 
relationship between the external momenta and the scale $\mu$;
for example, we are free to choose to place our external particles
on-shell as in Sect.~3, if desired.

Given the above effective field theories, we would then proceed
to compare with experiment in the usual way.
We would begin with our effective Lagrangian formulated in terms
of our KK parameters $\lambda$ evaluated for an arbitrary scale $\mu$:
\beq
            {\cal L}~=~ {\cal L}[ \lambda_i(\mu) ]~. 
\eeq
Since experimentalists will measure physical 
observables such as cross sections and decay rates
at a particular energy $E$, we would then calculate such quantities
from our effective Lagrangian.  Denoting such physical observables collectively
as $\sigma$, we would thereby obtain predicted values for these
quantities in terms of $\mu$, $E$, and our unknown parameters $\lambda_i(\mu)$:
\beq
             \sigma_{\rm pred}~=~
             \sigma_{\rm pred}[ \lambda_i(\mu), \mu, E]~.
\eeq
Setting $\sigma_{\rm pred}=\sigma_{\rm expt}$ then allows us to 
constrain the $\lambda_i(\mu)$ parameters, and thereby
deduce the effective Lagrangian experimentally.
Note, in this context,  that while the 
values of the $\lambda_i(\mu)$ parameters
in the effective Lagrangian
will depend on the specific chosen reference value of $\mu$,
the final result for the 
physically observable cross sections and decay rates 
are independent of $\mu$, as always.

\section{Conclusions and Future Directions
\label{conclude}
}
\setcounter{footnote}{0}

This paper concludes the two-part series initiated in Ref.~\cite{paper1a}.
In Ref.~\cite{paper1a},
        we proposed two new regulators 
        for quantum field theories in spacetimes with compactified extra dimensions.
        Unlike most other regulators which have been used in
        the extra-dimension literature, these regulators are 
        specifically designed to respect the original
        higher-dimensional Lorentz and gauge symmetries that exist prior to
        compactification, and not merely the four-dimensional symmetries which
        remain afterward.

In this paper, we continued this work by showing 
how these regulators may be used in order
        to extract ultraviolet-finite results from one-loop
        calculations.  
We provided a general procedure which accomplishes this, and
demonstrated its use through two explicit examples.
We also showed how this formalism allows us 
        to derive Wilsonian effective field theories for 
        Kaluza-Klein modes at different energy scales.  

The key property underpinning our methods
is that the divergent corrections to parameters describing the physics of the excited 
        Kaluza-Klein modes are absorbed into
        the corresponding parameters for zero modes.
This eliminates the need to
        introduce independent counterterms for parameters characterizing
        different Kaluza-Klein modes.  
        Our effective field theories can therefore simplify radiative calculations 
        involving towers Kaluza-Klein modes,
and should be especially relevant if data from the LHC should
happen to suggest the existence of TeV-sized extra dimensions.
Indeed, when the parameters describing the zero modes
are taken as experimental inputs, the relative corrections $\Delta(\lambda_n-\lambda_0)$
that we have determined will enable us to predict the properties
of the entire corresponding KK spectrum.
Knowledge of these differences thus allows
us to obtain regulator-independent (\ie, $\Lambda$-independent) EFT's for KK modes 
at various energy scales.

Despite the fact that our final results are regulator-independent,
as required,
we stress that the approaches we have taken here
rely rather crucially on the existence of such regulators in order
to perform the explicit calculations.
If we had not employed regulators which explicitly preserve the higher-dimensional
Lorentz and gauge symmetries that exist prior to compactification, 
we would have obtained additional spurious terms which would not have been
easy to disentangle from the physical effects of compactification.
Note that in some sense, all of the radiative corrections we have been calculating
represent the breaking of five-dimensional Lorentz invariance due to 
compactification:  this breaking, which originates {\it non-locally}\/
through the compactification of the spacetime geometry in the UV limit,
becomes a {\it local}\/ effect in the EFT at lower energy scales.
It would thus have been difficult to separate these 5D 
Lorentz-violating terms from those that would have
also emerged from a poor choice of UV regulator.  However, by employing
the regulators we developed in Ref.~\cite{paper1a}, we have avoided the
introduction of such spurious terms altogether.  The 5D Lorentz-violating effects
of the radiative corrections we have calculated can thus be interpreted
directly as the low-energy consequences of spacetime compactification, as expected. 

As already noted at the end of Ref.~\cite{paper1a}, our analysis
has been limited in a number of significant ways.
For example, this analysis has been restricted
to five dimensions and to one-loop amplitudes.
We have also focused on flat compactification spaces without
orbifold fixed points.   However,  compactification on orbifolded 
geometries is ultimately required in order to obtain a chiral theory 
in four dimensions.   Therefore, although our results can be taken to apply to the
bulk physics in such setups, they would require generalization before they could
accommodate orbifold fixed-point contributions such as those which
might emerge from, \eg,  brane-kinetic terms. 

Even within the framework of compactification of a single
extra dimension on a circle, there remain important extensions
which can also be considered.
For example, when deriving our effective field theories in Sect.~4,
we focused on the scale-dependence of those parameters $\lambda_i$ 
(\eg, KK masses and couplings) which
already appeared in the original UV Lagrangian.
However, as we pass to lower energies, 
new effective operators, \ie, new higher-order interactions, will be generated.
Despite the intrinsic non-renormalizability of higher-dimensional
theories, such operators have generally been ignored in the
higher-dimensional literature. 
Fortunately, our techniques have the advantage of being
able to handle these new interactions just as easily as
they handle the leading interactions we have already considered;
all that changes is the form of the $f$-functions.
Although the extra contributions from such operators
are generally suppressed compared with the leading
contributions which have been our focus, they can give rise to important 
phenomenological effects and thus must be taken into account in any complete
study of the low-energy phenomenologies of KK theories.

The methods developed in this paper open intriguing possibilities for future
phenomenological studies.  For example, in an upcoming paper~\cite{paper2},
we shall calculate KK radiative corrections and derive EFT's 
for two specific higher-dimensional models.
We shall also generally study how KK mass relations such
as $m_n^2=m_0^2 + n^2/R^2$ ``evolve'' as we pass from the UV limit
to the lower energy scales at which such KK states might eventually
be discovered.
Such analyses can thus be of direct importance for the discovery
and interpretation of such states in future collider experiments.

\bigskip
\eject

\section*{Acknowledgments}
\setcounter{footnote}{0}

We wish to thank 
Z.~Chacko and U.~van Kolck 
for discussions pertaining to effective field theories
in the context of extra dimensions.
This work is supported in part by the National Science Foundation
under Grant PHY/0301998, by the Department of Energy under
Grant~DE-FG02-04ER-41298, and by a Research Innovation Award from
Research Corporation.

\bigskip


\bibliographystyle{unsrt}
\section*{References}

\noindent {\it Note}\/:   This paper is the second in a two-part series.
      Complete references for this series are listed in
      Ref.~\protect\cite{paper1a}.

\end{document}